\documentclass[prd,twocolumn,showpacs,amsmath,amssymb,superscriptaddress,floatfix,nofootinbib]{revtex4-1}

\usepackage{amsmath, epsfig, amssymb, inputenc}
\usepackage[usenames]{color}
\usepackage{slashed}
\usepackage{multirow}
\usepackage{dcolumn}
\usepackage{subfigure}
\usepackage{dsfont}
\usepackage{comment}
\usepackage{steinmetz}

\allowdisplaybreaks

\usepackage{hyperref}
\hypersetup{
    colorlinks=true,
    citecolor=blue,
    linkcolor=blue,
    urlcolor=blue,
    pdfborder={0 0 0}
}

\begin{document}

\title{\texorpdfstring{Revisiting $B_c^{(*)}$ production in $Z$-boson decays at NLO in QCD}{Revisiting Bc and Bc* production in Z-boson decays at NLO in QCD}}

\author{Qi-Wei Ke}\email{keqw@cqu.edu.cn}
\affiliation{Department of Physics, Chongqing Key Laboratory for Strongly Coupled Physics, Chongqing University, Chongqing 401331, the People's Republic of China}
\author{Lei Guo}\email{guoleicqu@cqu.edu.cn}
\affiliation{Department of Physics, Chongqing Key Laboratory for Strongly Coupled Physics, Chongqing University, Chongqing 401331, the People's Republic of China}
\author{Jun Zeng}\email{zengj@hainnu.edu.cn}
\affiliation{College of Physics and Electronic Engineering, Hainan Normal University, Haikou 571158, Hainan, the People's Republic of China}
\author{Xu-Chang Zheng}\email{zhengxc@cqu.edu.cn}
\affiliation{Department of Physics, Chongqing Key Laboratory for Strongly Coupled Physics, Chongqing University, Chongqing 401331, the People's Republic of China}

\begin{abstract}
We study $B_c^{(*)}$ production in $Z$-boson decays at next-to-leading
order (NLO) in QCD within the color-singlet nonrelativistic QCD
framework. The analysis covers both the $1S$ and $2S$ pseudoscalar
and vector states, extending the phenomenological study to radially
excited $B_c$ mesons. We calculate direct production widths and
energy-fraction and invariant-mass distributions, including both
charm- and bottom-quark production topologies.
For the central inputs and $\mu_R=2m_b$, the $K$ factors are
$1.52$ for $B_c$ and $1.33$ for $B_c^*$.
We examine uncertainties from the renormalization scale, heavy-quark
masses, and wave functions at the origin, and apply the principle of
maximum conformality to fix the renormalization scale from the
$\beta_0$-dependent NLO contribution.
The NLO corrections enhance the small- and intermediate-energy-fraction
regions and modify the invariant-mass spectra. We discuss these
features in relation to heavy-quark fragmentation.
The results provide production-rate and kinematic inputs for studies
of ground-state and radially excited $B_c$ mesons at high-statistics
$Z$ factories.
\end{abstract}

\maketitle

\section{Introduction}
\label{secIntro}

High-statistics samples of $Z$ bosons offer opportunities to study the
production of heavy quarkonia and mesons containing two different heavy
flavors. Proposed $e^+e^-$ colliders with a $Z$-pole program, including
FCC-ee and CEPC \cite{FCC:2018byv,CEPCStudyGroup:2018ghi}, motivate
improved theoretical descriptions of these processes. The hierarchy
between $m_Z$ and the heavy-quark masses allows short-distance production
to be treated perturbatively, while bound-state formation requires
nonperturbative input. Studies of heavy-quarkonium production in $Z$
decays illustrate the role of QCD corrections in this separation
\cite{Braaten:1993mp,Li:2010xu,Zheng:2021jyd}.

The $B_c$ meson, first observed by the CDF Collaboration \cite{CDF:1998ihx,CDF:1998axz},
contains two different heavy flavors. Its production can be described
within nonrelativistic QCD (NRQCD) factorization
\cite{Bodwin:1994jh}, with heavy-quark masses and wave functions at the
origin providing essential inputs. Potential-model and lattice-QCD
studies constrain the spectrum and related bound-state properties
\cite{Eichten:1994gt,Eichten:1995ch,Eichten:2019gig,Gayer:2024akw}.
The observation of excited $B_c$ states by the ATLAS, CMS, and LHCb
Collaborations \cite{Aad:2014laa,CMS:2019bc2s,LHCb:2019bc2s} further motivates
studies of both ground-state and excited-state production.

The channels $Z\to B_c^{(*)}+b+\bar c$ have been studied at leading
order (LO) \cite{Chang:1992bb,Kiselev:1994qp,Deng:2010iq}.
NLO QCD calculations already exist for the pseudoscalar channel
\cite{Qiao:2011zc} and the vector channel \cite{Jiang:2015bcstar}.
Reference~\cite{Zheng:2017xgj} subsequently treated associated
$B_c^{(*)}$ production around the $Z$ pole and explicitly compared
the corresponding $Z$ decay widths with earlier calculations.
That comparison found agreement for the vector channel but a discrepancy
for the pseudoscalar channel. These calculations provide the basis for
further phenomenological studies of the $B_c$ family in $Z$ decays,
including excited-state production and differential observables.

Heavy-quark fragmentation provides a complementary description at large
energies. Both $\bar b$ and $c$ can fragment into a $c\bar b$ meson,
with different fragmentation probabilities
\cite{Braaten:1993jn,Zheng:2019gnb}.
The invariant masses of the meson with the accompanying heavy quarks
can help characterize these configurations. At finite energy, however,
the full amplitude contains interference and contributions beyond a
simple fragmentation approximation. A peak in a distribution therefore
does not by itself establish a unique production mechanism.

In this work, we extend the NLO phenomenological analysis to the $2S$
pseudoscalar and vector states and present a common treatment of
ground-state and radially excited $B_c^{(*)}$ production in $Z$ decays.
We calculate the integrated widths and the energy-fraction and
invariant-mass distributions for all four states. We examine their
dependence on the renormalization scale, heavy-quark masses, and
wave functions at the origin, and apply the principle of maximum
conformality (PMC) to absorb the $\beta_0$-dependent NLO term into
the running coupling.
These predictions are relevant to both direct excited-state production
and its contribution to the ground-state yield through feed-down.
We use $B_c^{(*)}$ for the $c\bar b$ state in the amplitudes. Its
charge-conjugate channel is distinct and has the same width at the
accuracy considered; a sum over both charges requires an explicit
factor of two.

Section~\ref{widthLO} defines the LO amplitudes and normalization.
Section~\ref{secNLO} describes the NLO corrections.
Section~\ref{secNumer} presents the numerical results, and
Sec.~\ref{secSum} summarizes our findings.

\section{LO decay width}
\label{widthLO}

Within NRQCD factorization, the differential width for producing a
physical meson $H=B_c$ or $B_c^*$ is written as
\begin{equation}
d\Gamma_H=\sum_n d\widetilde{\Gamma}_{(c\bar b)[n]}
                 \langle{\cal O}^{H}(n)\rangle .
\label{nrqcdfact}
\end{equation}
Here $d\widetilde{\Gamma}$ is a short-distance coefficient (SDC) and
$\langle{\cal O}^{H}(n)\rangle$ is a long-distance matrix element (LDME).
The label $n$ specifies the spin, orbital angular momentum, and color
state of the pair. We retain the leading color-singlet channels
$^1S_0^{[1]}$ and $^3S_1^{[1]}$.

We assign the momenta as
$Z(p_0)\to H(p_1)+b(p_2)+\bar c(p_3)$ and define $M=m_b+m_c$.
The constituent momenta are
\begin{equation}
p_c=\frac{m_c}{M}p_1+q,\qquad
p_{\bar b}=\frac{m_b}{M}p_1-q .
\end{equation}
At leading order in the relative velocity, $q=0$ and $p_1^2=M^2$.
The same mass $M$ must be used in the constituent momenta, phase space,
and polarization sum. Spin splittings and radial-excitation mass
differences are beyond this approximation.

For the explicit amplitudes below, we absorb the wave function into a
physical-state projector \cite{Jiang:2015bcstar,Zheng:2017xgj}:
\begin{align}
v_j(p_{\bar b})\bar u_k(p_c)&\longrightarrow
             \frac{\delta_{jk}}{\sqrt{N_c}}\,{\cal P}_H,\nonumber\\
{\cal P}_H&=\frac{\Psi_H(0)}{2\sqrt M}\,
                  \Omega_H(\slashed p_1+M), \label{eq.projector}\\
\Omega_{B_c}&=\gamma_5,\qquad
\Omega_{B_c^*}=\slashed\epsilon_H^*,\qquad
\Psi_H(0)=\frac{R_H(0)}{\sqrt{4\pi}}.\nonumber
\end{align}
An overall phase of the projector does not affect the width.
In the conventional spin-summed LDME normalization,
\begin{equation}
\langle{\cal O}_1^H({}^{2S+1}S_J)\rangle
=2N_c(2J+1)|\Psi_H(0)|^2[1+{\cal O}(v^2)].
\end{equation}
The amplitudes written with Eq.~(\ref{eq.projector}) already include the
bound-state normalization and must not be multiplied by this LDME again.
Alternatively, an SDC can be defined by removing the corresponding LDME
from the physical width in Eq.~(\ref{nrqcdfact}).

The vector-state polarization sum is
\begin{equation}
\sum_\lambda\epsilon_{H\alpha}(p_1,\lambda)
 \epsilon_{H\beta}^*(p_1,\lambda)
=-g_{\alpha\beta}+\frac{p_{1\alpha}p_{1\beta}}{M^2}.
\end{equation}

The Born contribution is
\begin{equation}
d\Gamma_{\rm LO}^{H}
=\frac{1}{2m_Z}\frac{1}{3}
  \sum|\mathcal{M}_{\rm LO}|^2d\Phi_3 .
\end{equation}
The sum runs over final-state spins and colors and the three physical
$Z$ polarizations; the initial-state average is supplied only by the
explicit factor $1/3$. In dimensional regularization,
\begin{equation}
d\Phi_3=(2\pi)^d\delta^{(d)}\!
 \left(p_0-\sum_{f=1}^{3}p_f\right)
 \prod_{f=1}^{3}\frac{d^{d-1}{\bf p}_f}{(2\pi)^{d-1}2E_f},
\label{dphi3}
\end{equation}
with $d=4-2\epsilon$.

\begin{figure}[htbp]
    \centering
    \includegraphics[width=0.9\linewidth]{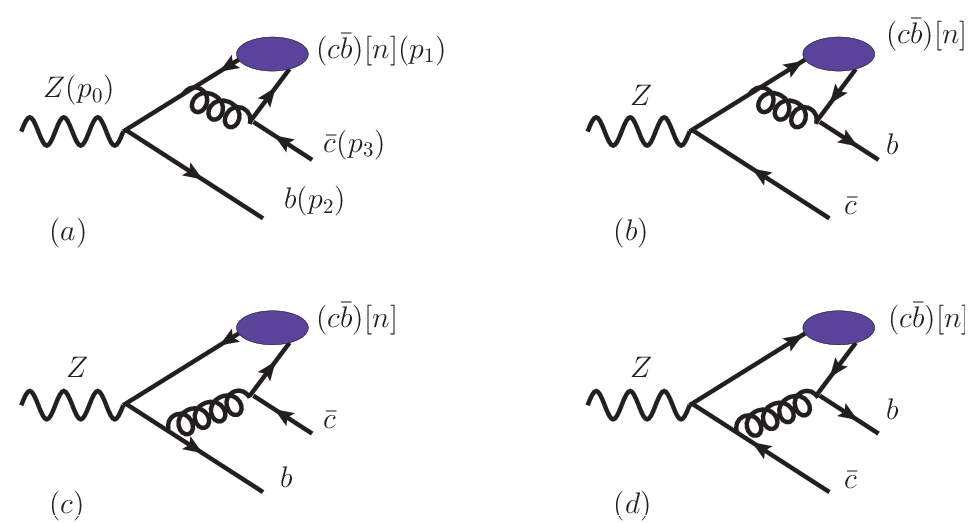}
    \caption{Tree-level diagrams for $Z(p_0)\to H(p_1)+b(p_2)+\bar c(p_3)$,
    where $H$ is a color-singlet $c\bar b$ state. Diagrams (a) and (c)
    contain the $Zb\bar b$ vertex, while (b) and (d) contain the
    $Zc\bar c$ vertex.}
    \label{feylo}
\end{figure}

The four tree-level diagrams are shown in Fig.~\ref{feylo}.
We label their Dirac chains as $\mathcal A_a$, $\mathcal A_b$,
$\mathcal A_c$, and $\mathcal A_d$, following panels (a)--(d).
The gluon momenta and heavy-quark propagators are defined by
\begin{align}
k_c&=p_c+p_3,\qquad k_b=p_{\bar b}+p_2,\nonumber\\
S_Q(\ell)&=\frac{\slashed\ell+m_Q}{\ell^2-m_Q^2+i0},\qquad
{\cal V}_Q=\slashed\epsilon_Z(V_Q-A_Q\gamma_5).
\end{align}
We use $g=e/\sin\theta_W$ and
\begin{align}
V_c&=\frac12-\frac43\sin^2\theta_W,& A_c&=\frac12,\nonumber\\
V_b&=-\frac12+\frac23\sin^2\theta_W,& A_b&=-\frac12.
\end{align}
The color-singlet contraction is performed once:
\begin{equation}
\sum_{a,j,k}T^a_{ij}\frac{\delta_{jk}}{\sqrt{N_c}}T^a_{kl}
 =\frac{C_F}{\sqrt{N_c}}\delta_{il}.
\label{eq.color}
\end{equation}
With this color factor, the sum of the four diagrams in
Fig.~\ref{feylo} gives the Born amplitude, up to a common overall phase,
\begin{align}
\mathcal{M}_{\rm LO}
&={\cal C}_{il}\bar u_i(p_2)
 \left[\frac{{\cal A}_a+{\cal A}_c}{k_c^2+i0}
       +\frac{{\cal A}_b+{\cal A}_d}{k_b^2+i0}\right]v_l(p_3),
 \nonumber\\
{\cal C}_{il}&=\frac{g g_s^2}{2\cos\theta_W}
                 \frac{C_F}{\sqrt{N_c}}\delta_{il},
\label{eq.born}
\end{align}
where repeated external color labels are contracted. The Dirac chains
for Fig.~\ref{feylo}(a)--(d) are
\begin{equation}
\begin{aligned}
{\cal A}_a&={\cal V}_b S_b(-p_1-p_3)
                  \gamma^\mu{\cal P}_H\gamma_\mu,\\
{\cal A}_b&=\gamma^\mu{\cal P}_H\gamma_\mu
                  S_c(p_1+p_2){\cal V}_c,\\
{\cal A}_c&=\gamma^\mu S_b(p_0-p_{\bar b})
                  {\cal V}_b{\cal P}_H\gamma_\mu,\\
{\cal A}_d&=\gamma^\mu{\cal P}_H{\cal V}_c
                  S_c(p_c-p_0)\gamma_\mu .
\end{aligned}
\label{eq.bornchains}
\end{equation}
The first and third diagrams involve the $Zb\bar b$ coupling and
the gluon momentum $k_c$, while the second and fourth involve the
$Zc\bar c$ coupling and the gluon momentum $k_b$.
The internal momenta follow from $p_0=p_1+p_2+p_3$.

\section{NLO corrections}
\label{secNLO}

The NLO QCD correction consists of virtual one-loop corrections to the
three-body channel and real-gluon emission in $Z\to H+b+\bar c+g$.
Ultraviolet (UV) and infrared (IR) singularities are regularized in
$d=4-2\epsilon$ dimensions. The treatment of $\gamma_5$ is required
both for the axial $Z$ coupling and for the pseudoscalar projector.
We adopt the anticommuting-$\gamma_5$ prescription of
Ref.~\cite{Korner:1991sx}, with a fixed reading point for traces
containing $\gamma_5$. In this prescription, $\gamma_5$ anticommutes
with the Dirac matrices in $d$ dimensions, while trace cyclicity is
not used for traces containing a single $\gamma_5$.

\subsection{Virtual NLO correction}

\begin{figure}[htbp]
    \centering
    \includegraphics[width=0.9\linewidth]{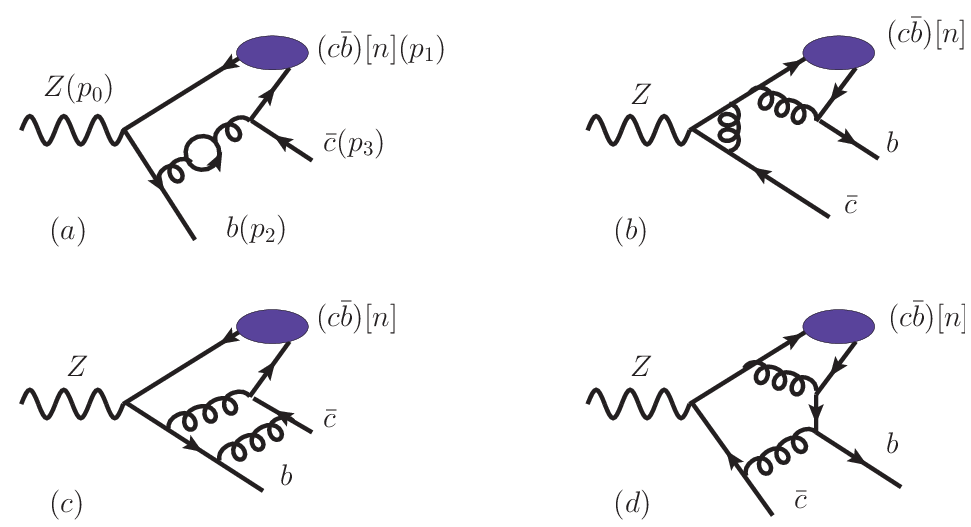}
    \caption{Four typical one-loop Feynman diagrams for $Z \to (c\bar{b})+b+\bar{c}$.}
    \label{feyvir}
\end{figure}

The virtual term is evaluated by interfering the renormalized one-loop
amplitude with the Born amplitude.  Four typical one-loop Feynman diagrams are shown in Fig.~\ref{feyvir}.  Its contribution to the
differential width is
\begin{equation}
d\Gamma_{\rm Virtual}^{H}=\frac{1}{3}\frac{1}{2m_Z}\sum 2 {\rm Re}\left(\mathcal{M}^*_{\rm LO} \mathcal{M}_{\rm Virtual} \right)d\Phi_3,
\end{equation}
where $\mathcal{M}_{\rm Virtual}$ is the renormalized one-loop amplitude and the
three-body phase space is the one defined in Eq.~(\ref{dphi3}).

For the color-singlet $S$-wave channels, the short-distance matching is
performed by expanding the amplitude in the relative momentum of the
constituent heavy quarks.  Since only the leading term in the velocity
expansion is needed, $q$ is set to zero before the loop integration.  In the
method-of-regions interpretation \cite{Beneke:1997zp}, this procedure selects
the hard part of the partonic amplitude; the Coulomb contribution is assigned
to the LDME and is not kept as a separate perturbative term.

The loop amplitudes contain both UV and IR poles.  The former are subtracted
by renormalization, while the latter cancel only after the real-emission
channel is included.  We renormalize the heavy-quark fields, heavy-quark
masses, and gluon field in the on-mass-shell (OS) scheme, and define the
strong coupling in the $\overline{\rm MS}$ scheme. To express the
renormalization constants $\delta Z_i\equiv Z_i-1$
\cite{Klasen:2004tz} compactly, we define
\begin{equation}
\Delta_{\rm UV,IR}=\frac{1}{\epsilon_{\rm UV,IR}}
                   -\gamma_E+\ln(4\pi).
\end{equation}
The counterterms are
\begin{align}
\delta Z_{2}^{\rm OS}
 &=-C_F\frac{\alpha_s}{4\pi}
   \left[\Delta_{\rm UV}+2\Delta_{\rm IR}
          +3\ln\frac{\mu_R^2}{m^2}+4\right],\nonumber\\
\delta Z_m^{\rm OS}
 &=-3C_F\frac{\alpha_s}{4\pi}
   \left[\Delta_{\rm UV}+\ln\frac{\mu_R^2}{m^2}+\frac43\right],\nonumber\\
\delta Z_3^{\rm OS}
 &=\frac{\alpha_s}{4\pi}
  \left[(\beta'_0-2C_A)(\Delta_{\rm UV}-\Delta_{\rm IR})\right.\nonumber\\
 &\hspace{1.5em}\left.
   -\frac43T_F\sum_{Q\in{\cal Q}_h}
    \left(\Delta_{\rm UV}+\ln\frac{\mu_R^2}{m_Q^2}\right)\right],\nonumber\\
\delta Z_g^{\overline{\rm MS}}
 &=-\frac{\beta_0^{(n_f)}}{2}\frac{\alpha_s}{4\pi}\Delta_{\rm UV}.
\end{align}
Here $m$ in $\delta Z_{2}^{\rm OS}$ and $\delta Z_m^{\rm OS}$ is the mass
of the external heavy quark, and $\mu_R$ and $\gamma_E$ denote the
renormalization scale and Euler constant. The coefficients of the
QCD $\beta$ function are
\begin{equation}
\beta_0^{(n_f)}=\frac{11}{3}C_A-\frac{4}{3}T_F n_f,\qquad
\beta'_0=\frac{11}{3}C_A-\frac{4}{3}T_F n_{lf},
\end{equation}
where $n_{lf}=3$, $C_A=3$, $C_F=4/3$, and $T_F=1/2$.
We use $n_f=4$ for $m_c\leq\mu_R<m_b$ and $n_f=5$ for
$m_b\leq\mu_R<m_t$. Below the bottom threshold, the bottom-quark
vacuum-polarization contribution is decoupled, and charm is the
only active massive flavor, so that $\mathcal{Q}_h=\{c\}$.
Above the threshold, both charm and bottom are active, with
$\mathcal{Q}_h=\{c,b\}$. The same active-flavor prescription is used
in the renormalization constants and the running of $\alpha_s$.
The top-quark vacuum-polarization contribution is neglected.

The algebraic calculation is organized as follows.  FeynArts
\cite{Hahn:2000kx} is used to generate the diagrams and amplitudes.
FeynCalc \cite{Mertig:1990an,Shtabovenko:2016sxi} performs the color and
Dirac algebra.  \texttt{Apart} \cite{Feng:2012iq} is used to separate
linearly dependent denominators, and FIRE \cite{Smirnov:2008iw} reduces the
loop integrals to master integrals, which are evaluated numerically with
LoopTools \cite{Hahn:1998yk}.  The phase-space integrations are carried out
with Vegas \cite{Lepage:1977sw}.  As internal checks, we verify the
cancellation of UV poles after renormalization and the cancellation of IR
poles after combining the virtual and real pieces.

\subsection{Real NLO correction}

\begin{figure}[htbp]
    \centering
    \includegraphics[width=0.9\linewidth]{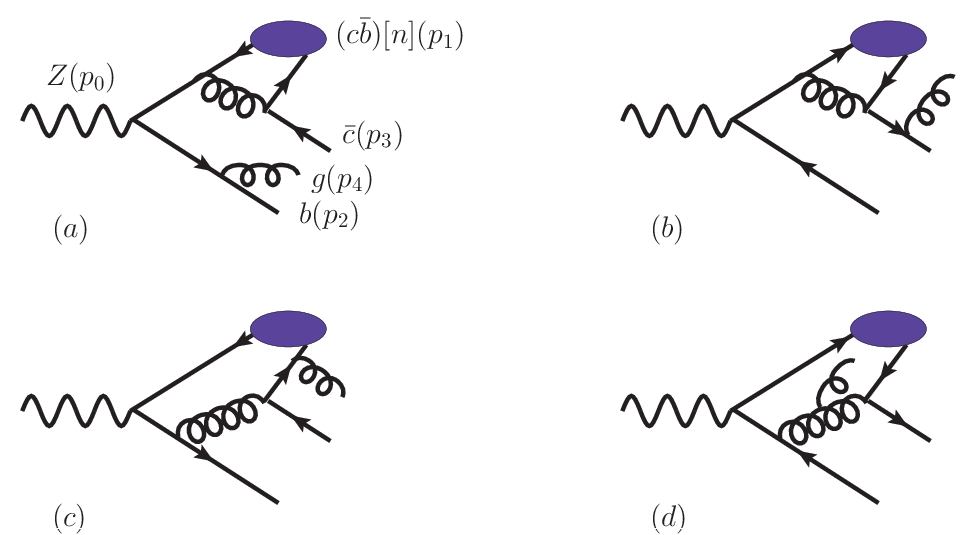}
    \caption{Four typical real-correction Feynman diagrams for $Z \to (c\bar{b})+b+\bar{c}+g$.}
    \label{feyreal}
\end{figure}

The real-emission part comes from the four-body channel
$Z(p_0) \to (c\bar{b})(p_1)+b(p_2)+\bar{c}(p_3)+g(p_4)$.
Four typical real-emission diagrams are displayed in Fig.~\ref{feyreal}.

Denoting the corresponding amplitude by $\mathcal{M}_{\rm Real}$, we write
\begin{eqnarray}
d\Gamma_{\rm Real}^{H}=\frac{1}{3}\frac{1}{2m_Z}\sum \vert \mathcal{M}_{\rm Real} \vert^2 d\Phi_4,
\end{eqnarray}
where the four-body phase space is
\begin{eqnarray}
d\Phi_4=(2\pi)^d\delta^{(d)}\left(p_0-\sum_{f=1}^4 p_f\right)\prod_{f=1}^4 \frac{d^{d-1} \textbf{p}_f}{(2\pi)^{d-1} 2E_f}.
\end{eqnarray}

The only infrared singularity in the real channel is the soft singularity
associated with $p_4\to0$; the finite charm- and bottom-quark masses regulate
the collinear configurations.  It is therefore sufficient to introduce a
single slicing parameter $\delta_s$.  In the $Z$ rest frame, the gluon-energy
regions are defined by $E_4 \leq m_Z\delta_s/2$ and
$E_4>m_Z\delta_s/2$, leading to
\begin{eqnarray}
d\Gamma_{\rm Real}^{H}=d\Gamma_{\rm S}^{H}+d\Gamma_{\rm H}^{H},
\end{eqnarray}
where
\begin{eqnarray}
d\Gamma_{\rm S}^{H}=\frac{1}{3}\frac{1}{2m_Z}\sum \vert \mathcal{M}_{\rm Real} \vert^2 d\Phi_4 \vert_{E_4 \leq m_Z \delta_s/2},
\label{eq.gamma-soft}
\end{eqnarray}
and
\begin{eqnarray}
d\Gamma_{\rm H}^{H}=\frac{1}{3}\frac{1}{2m_Z}\sum \vert \mathcal{M}_{\rm Real} \vert^2 d\Phi_4 \vert_{E_4 > m_Z \delta_s/2}.
\end{eqnarray}

In the soft domain, the squared amplitude factorizes into the Born term times
the eikonal factor \cite{Harris:2001sx,Bassetto:1983mvz}:
\begin{eqnarray}
&&\sum \vert \mathcal{M}_{\rm Real} \vert^2 \vert_{E_4 \leq m_Z \delta_s/2} \nonumber \\
&&= 4\pi \alpha_s C_F \mu_R^{2\epsilon}\bigg[ -\frac{p_2^2}{(p_2\cdot p_4)^2}+\frac{2p_2\cdot p_3}{(p_2\cdot p_4)(p_3 \cdot p_4)} \nonumber \\
&&~~~ -\frac{p_3^2}{(p_3\cdot p_4)^2}\bigg]\sum \vert \mathcal{M}_{\rm LO} \vert^2.
\label{eq.amp-soft}
\end{eqnarray}
At $q=0$, the two constituents of the color-singlet pair have the same
four-velocity. Their leading eikonal color charges cancel, leaving the
dipole formed by the open $b$ and $\bar c$ lines in
Eq.~(\ref{eq.amp-soft}).
The same soft limit factorizes the four-particle phase space, up to terms
vanishing with $\delta_s\to 0$ \cite{Harris:2001sx}:
\begin{eqnarray}
d\Phi_4 \vert_{E_4 \leq m_Z \delta_s/2}=d\Phi_3 \frac{d^{d-1} \textbf{p}_4}{(2\pi)^{d-1} 2E_4}\vert_{E_4 \leq m_Z \delta_s/2},
\label{eq.phs-soft}
\end{eqnarray}
where $d\Phi_3$ is the underlying three-body phase space in
Eq.~(\ref{dphi3}).

Combining Eqs.~(\ref{eq.amp-soft}) and (\ref{eq.phs-soft}) with
Eq.~(\ref{eq.gamma-soft}), the analytical integration over the soft-gluon
momentum gives \cite{Denner:1991kt,Beenakker:2002nc}
\begingroup
\small
\begin{eqnarray}
&&d\Gamma^{H}_{\rm S}\nonumber\\
&&=d\Gamma^{H}_{\rm LO}\left[\frac{C_F \alpha_s \Gamma(1+\epsilon)}{\pi}\left(\frac{4\pi\mu_R^2}{m_Z^2}\right)^{\epsilon}\right]
\Bigg\{\left(\frac{1}{\epsilon}-\ln\delta_s^2\right)\nonumber\\
&&\;\times\left(1-\frac{\kappa\,p_2\cdot p_3}{\kappa^2m_b^2-m_c^2}\ln\frac{\kappa^2m_b^2}{m_c^2}\right)
+\frac{1}{2\beta_2}\ln\frac{1+\beta_2}{1-\beta_2}\nonumber\\
&&\;+\frac{1}{2\beta_3}\ln\frac{1+\beta_3}{1-\beta_3}
+\frac{2\kappa\,p_2\cdot p_3}{\kappa^2m_b^2-m_c^2}
\Bigg[\frac14\ln^2\frac{u^0-|\mathbf u|}{u^0+|\mathbf u|}\nonumber\\
&&\;\quad+{\rm Li}_2\left(1-\frac{u^0+|\mathbf u|}{v}\right)\nonumber\\
&&\;\quad+{\rm Li}_2\left(1-\frac{u^0-|\mathbf u|}{v}\right)
\Bigg]\Bigg\vert_{u=p_3}^{u=\kappa p_2}\Bigg\},
\end{eqnarray}
\endgroup
where
\begin{eqnarray}
\beta_2&=&\sqrt{1-m_b^2/E_2^2},\nonumber \\
\beta_3&=&\sqrt{1-m_c^2/E_3^2},\nonumber \\
v&=& \frac{\kappa^2m_b^2-m_c^2}{2(\kappa\, E_2-E_3)},\nonumber \\
\kappa&=& \frac{p_2\cdot p_3+\sqrt{(p_2 \cdot p_3)^2-m_b^2m_c^2}}{m_b^2}, 
\end{eqnarray}
with $E_2$ and $E_3$ evaluated in the $Z$ rest frame.

The hard remainder is free of singularities and is integrated directly in
four dimensions. The dependence on $\delta_s$ cancels between the soft
and hard terms up to the power corrections inherent in phase-space
slicing. In the numerical calculation, $\delta_s$ is varied between
$10^{-5}$ and $10^{-7}$ to check the stability of the result.

Combining the renormalized virtual and real contributions gives the
finite NLO width. With the physical-state projector of
Eq.~(\ref{eq.projector}), the wave function is already included in
each amplitude. An additional multiplication by the LDME would
double count the bound-state normalization. In an SDC-based
implementation, the equivalent result follows by inserting the
consistently normalized SDC and LDME in Eq.~(\ref{nrqcdfact}).

\section{Numerical results and discussion}
\label{secNumer}

We use the following benchmark inputs. The heavy-quark masses are
treated as pole-mass parameters, and their variation ranges are
sensitivity estimates rather than statistical confidence intervals.
The electroweak inputs follow Ref.~\cite{Workman:2022ynf}:
\begin{equation}
\begin{aligned}
m_c&=1.5\pm0.1\,{\rm GeV},\quad m_b=4.9\pm0.2\,{\rm GeV},\\
m_Z&=91.1876\,{\rm GeV},\quad \sin^2\theta_W=0.231,\\
\alpha&=1/128,
\end{aligned}
\label{eq.parameter}
\end{equation}
where $m_c$ and $m_b$ are pole masses and $\alpha$ is the electromagnetic
coupling constant at the $Z$-boson scale.  The scale evolution of the strong
coupling is implemented with the two-loop expression

\begin{equation}
\alpha_s(\mu_R)=\frac{4\pi}{\beta_0 L_R}
  \left[1-\frac{\beta_1\ln L_R}{\beta_0^2L_R}\right],
\qquad L_R=\ln\frac{\mu_R^2}{\Lambda_{\rm QCD}^2},
\label{eq.alphas}
\end{equation}
where $\beta_1=34C_A^2/3-4T_FC_Fn_f-20T_FC_An_f/3$.
Following the active-flavor prescription in Sec.~\ref{secNLO},
we use $n_f=4$ for $m_c\leq\mu_R<m_b$ and $n_f=5$ for
$m_b\leq\mu_R<m_t$.
In the five-flavor region, the benchmark value
$\alpha_s(m_Z)=0.1185$ \cite{Workman:2022ynf} corresponds to
$\Lambda_{\rm QCD}^{n_f=5}=0.233\,{\rm GeV}$.
For the four-flavor region below the bottom threshold, we use
$\Lambda_{\rm QCD}^{n_f=4}=0.337\,{\rm GeV}$.
The two flavor regions are connected by threshold matching at
$\mu_R=m_b$, with the renormalization constants treated consistently
according to the active-flavor prescription in Sec.~\ref{secNLO}.

The radial wave functions at the origin are listed in
Table~\ref{tab:rwf}. The QCD column denotes the Buchm\"uller--Tye
potential-model input used as the central choice.

\begin{table}[!htbp]
\centering
\caption{Squared radial wave functions at the origin, $|R_{bc}(0)|^2$ (in ${\rm GeV}^3$), for $c\bar b$ mesons in three potential models.}
\label{tab:rwf}
\small
\renewcommand{\arraystretch}{1.15}
\begin{tabular*}{\columnwidth}[t]{@{\extracolsep{\fill}}lccc@{}}
\toprule
Level & Logarithmic & QCD & Cornell (2019)\\
 & \cite{Eichten:1994gt} & \cite{Eichten:1994gt} & \cite{Eichten:2019gig}\\
\colrule
$1S$ & 1.508 & 1.642 & 1.994\\
$2S$ & 0.770 & 0.983 & 1.144\\
\botrule
\end{tabular*}
\end{table}

At this accuracy, the $1S$ and $2S$ states use the same mass $M=m_b+m_c$
and the same SDC for a given spin channel. Their spectra therefore obey
\begin{equation}
\frac{d\Gamma_{2S}}{dx}
=\frac{|R_{2S}(0)|^2}{|R_{1S}(0)|^2}
  \frac{d\Gamma_{1S}}{dx},
\label{eq.radial}
\end{equation}
where $x$ denotes any of the observables defined below.
The $2S$ result is a normalization rescaling within this approximation;
it does not include excitation-dependent phase space or relativistic
effects.

The widths below describe direct production. An inclusive ground-state
yield also receives feed-down contributions, schematically
\begin{equation}
\Gamma_{\rm incl}(B_c)=\Gamma_{\rm dir}(B_c)
+\sum_{H\ne B_c}\Gamma_{\rm dir}(H)\,
 {\cal B}(H\to B_c+X).
\end{equation}
The branching fractions here include the relevant decay chains;
the differential distributions also require their decay kinematics.
For a sample of $N_Z$ decays, the direct production yield is
$N_H=N_Z\Gamma_{\rm dir}(H)/\Gamma_Z$, with the same charge convention
used for $N_H$ and $\Gamma_{\rm dir}(H)$. A reconstructed yield further
requires decay branching fractions, acceptance, and efficiency.

\subsection{Integrated decay widths}

We first discuss the integrated widths for
$Z\to B_c+b+\bar c+X$ and $Z\to B_c^*+b+\bar c+X$, including the NLO QCD
correction.

\begin{table}[!htbp]
\centering
\caption{LO and NLO decay widths (in keV) and $K$ factors for
$Z\to B_c+b+\bar{c}+X$ at three renormalization scales.}
\label{tb.widthBc}
\small
\setlength{\tabcolsep}{3pt}
\renewcommand{\arraystretch}{1.15}
\begin{tabular*}{\columnwidth}[t]{@{\extracolsep{\fill}}cccD{.}{.}{3.3}D{.}{.}{3.3}D{.}{.}{1.3}@{}}
\toprule
$\mu_R$ & $\alpha_s(\mu_R)$ & Level & \multicolumn{1}{c}{$\Gamma_{\rm LO}$} & \multicolumn{1}{c}{$\Gamma_{\rm NLO}$} & \multicolumn{1}{c}{$K$}\\
\colrule
\multirow{2}{*}{$2m_c$} & \multirow{2}{*}{0.259} & $1^1S_0$ & 136.9 & 135.8 & 0.99\\
 & & $2^1S_0$ & 82.0 & 81.3 & 0.99\\[3pt]
\multirow{2}{*}{$2m_b$} & \multirow{2}{*}{0.180} & $1^1S_0$ & 66.1 & 100.4 & 1.52\\
 & & $2^1S_0$ & 39.6 & 60.1 & 1.52\\[3pt]
\multirow{2}{*}{$m_Z/2$} & \multirow{2}{*}{0.132} & $1^1S_0$ & 35.6 & 66.7 & 1.87\\
 & & $2^1S_0$ & 21.3 & 39.9 & 1.87\\
\botrule
\end{tabular*}
\end{table}
\begin{table}[!htbp]
\centering
\caption{LO and NLO decay widths (in keV) and $K$ factors for
$Z\to B_c^*+b+\bar{c}+X$ at three renormalization scales.}
\label{tb.widthBc*}
\small
\setlength{\tabcolsep}{3pt}
\renewcommand{\arraystretch}{1.15}
\begin{tabular*}{\columnwidth}[t]{@{\extracolsep{\fill}}cccD{.}{.}{3.3}D{.}{.}{3.3}D{.}{.}{1.3}@{}}
\toprule
$\mu_R$ & $\alpha_s(\mu_R)$ & Level & \multicolumn{1}{c}{$\Gamma_{\rm LO}$} & \multicolumn{1}{c}{$\Gamma_{\rm NLO}$} & \multicolumn{1}{c}{$K$}\\
\colrule
\multirow{2}{*}{$2m_c$} & \multirow{2}{*}{0.259} & $1^3S_1$ & 191.5 & 139.6 & 0.73\\
 & & $2^3S_1$ & 114.7 & 83.6 & 0.73\\[3pt]
\multirow{2}{*}{$2m_b$} & \multirow{2}{*}{0.180} & $1^3S_1$ & 92.8 & 123.7 & 1.33\\
 & & $2^3S_1$ & 55.6 & 74.0 & 1.33\\[3pt]
\multirow{2}{*}{$m_Z/2$} & \multirow{2}{*}{0.132} & $1^3S_1$ & 49.7 & 86.5 & 1.74\\
 & & $2^3S_1$ & 29.8 & 51.8 & 1.74\\
\botrule
\end{tabular*}
\end{table}

Tables~\ref{tb.widthBc} and \ref{tb.widthBc*} give the results obtained with
$m_c=1.5\,{\rm GeV}$, $m_b=4.9\,{\rm GeV}$, and the QCD-potential wave
functions in Table~\ref{tab:rwf}.  Here $\Gamma_{\rm NLO}$ denotes the sum of
the $\mathcal O(\alpha_s^2)$ Born term and the $\mathcal O(\alpha_s^3)$ correction, and
$K=\Gamma_{\rm NLO}/\Gamma_{\rm LO}$.  The three scale choices
$\mu_R=2m_c$, $2m_b$, and $m_Z/2$ illustrate the sizable residual scale
dependence at fixed order.  In the pseudoscalar channel the $K$ factor changes
from $0.99$ to $1.87$ across this range, while in the vector channel it
changes from $0.73$ to $1.74$.  Therefore the radiative correction cannot be
represented by a single universal rescaling of the LO result.  The equal $K$
factors for the $1S$ and $2S$ entries within each spin channel reflect the
leading nonrelativistic approximation, where the radial excitation changes the
LDME normalization but not the perturbative SDC.

For comparison with previous calculations, the ground-state NLO
widths at $\mu_R=2m_b$ and $m_Z/2$ agree at the sub-percent level
with the widths inferred from the $Z$-pole cross sections in
Ref.~\cite{Zheng:2017xgj}, using the same inputs.
In particular, the pseudoscalar result supports the larger NLO width
reported there in the comparison with Ref.~\cite{Qiao:2011zc}.
The vector-channel $K$ factor at $\mu_R=2m_b$ is also consistent
with Ref.~\cite{Jiang:2015bcstar}.

We assess renormalization-scale dependence and variations of the
heavy-quark masses and wave functions separately.
The Monte Carlo integration error in the total width is of
order $10^{-1}\,{\rm keV}$; it does not determine the precision of
individual differential bins.
We vary $\mu_R$ over $[2m_c,m_Z/2]$ with central scale $2m_b$, take
$m_b=4.9\pm0.2\,{\rm GeV}$ and $m_c=1.5\pm0.1\,{\rm GeV}$,
and use the logarithmic and Cornell entries in Table~\ref{tab:rwf}
to bracket the wave-function input. This scale interval is wider than
the commonly used factor-of-two interval about the central scale.
The separate variations give
\begin{widetext}
\begingroup
\small
\begin{eqnarray}
\Gamma^{\rm LO}_{Z \to B_c(1^1S_0)+b\bar{c}X} &=& 66.1^{+70.8+1.9+15.0+14.2}_{-30.6-1.4-11.5-5.4}\,{\rm keV},\nonumber \\
\Gamma^{\rm NLO}_{Z \to B_c(1^1S_0)+b\bar{c}X} &=& 100.4^{+35.4+2.7+23.5+21.5}_{-33.7-1.9-17.9-8.2}\,{\rm keV},\nonumber \\
\Gamma^{\rm LO}_{Z \to B_c(2^1S_0)+b\bar{c}X} &=& 39.6^{+42.4+1.1+9.0+6.5}_{-18.3-0.8-6.9-8.6}\,{\rm keV},\nonumber \\
\Gamma^{\rm NLO}_{Z \to B_c(2^1S_0)+b\bar{c}X} &=& 60.1^{+21.2+1.6+14.1+9.8}_{-20.2-1.1-10.7-13.0}\,{\rm keV},
\label{eq.uncert1}
\end{eqnarray}
and
\begin{eqnarray}
\Gamma^{\rm LO}_{Z \to B_c^*(1^3S_1)+b\bar{c}X} &=& 92.8^{+98.7+1.0+23.3+20.0}_{-43.1-1.1-18.0-7.6}\,{\rm keV},\nonumber \\
\Gamma^{\rm NLO}_{Z \to B_c^*(1^3S_1)+b\bar{c}X} &=& 123.7^{+15.9+0.4+32.8+26.5}_{-37.2-0.6-25.2-10.1}\,{\rm keV},\nonumber \\
\Gamma^{\rm LO}_{Z \to B_c^*(2^3S_1)+b\bar{c}X} &=& 55.6^{+59.1+0.6+13.9+9.1}_{-25.8-0.6-10.8-12.0}\,{\rm keV},\nonumber \\
\Gamma^{\rm NLO}_{Z \to B_c^*(2^3S_1)+b\bar{c}X} &=& 74.0^{+9.5+0.2+19.6+12.1}_{-22.3-0.3-15.1-16.0}\,{\rm keV}.
\label{eq.uncert2}
\end{eqnarray}
\endgroup
\end{widetext}

In each line, the uncertainties are ordered as
$\mu_R$, $m_b$, $m_c$, and $|R_{bc}(0)|^2$.
Scale dependence is substantial, but it is not the largest contribution
in every channel and direction. In particular, the charm-mass and
wave-function variations exceed the upward scale variation for the
NLO vector channel.

For an illustrative combined estimate, the four contributions are
treated as independent and added in quadrature. This convention does
not define a statistical confidence interval: scale and model
variations are not random measurement errors, and correlations
between masses and potential-model wave functions are not included.
Adding these contributions in quadrature gives

\begin{widetext}
\begin{eqnarray}
\Gamma^{\rm LO}_{Z \to B_c(1^1S_0)+b\bar{c}X} &=& 66.1^{+73.8}_{-33.1}\,{\rm keV},\nonumber \\
\Gamma^{\rm NLO}_{Z \to B_c(1^1S_0)+b\bar{c}X} &=& 100.4^{+47.7}_{-39.1}\,{\rm keV},\nonumber \\
\Gamma^{\rm LO}_{Z \to B_c(2^1S_0)+b\bar{c}X} &=& 39.6^{+43.8}_{-21.4}\,{\rm keV},\nonumber \\
\Gamma^{\rm NLO}_{Z \to B_c(2^1S_0)+b\bar{c}X} &=& 60.1^{+27.3}_{-26.3}\,{\rm keV},
\label{eq.uncert3}
\end{eqnarray}
and
\begin{eqnarray}
\Gamma^{\rm LO}_{Z \to B_c^*(1^3S_1)+b\bar{c}X} &=& 92.8^{+103.3}_{-47.3}\,{\rm keV},\nonumber \\
\Gamma^{\rm NLO}_{Z \to B_c^*(1^3S_1)+b\bar{c}X} &=& 123.7^{+45.1}_{-46.0}\,{\rm keV},\nonumber \\
\Gamma^{\rm LO}_{Z \to B_c^*(2^3S_1)+b\bar{c}X} &=& 55.6^{+61.4}_{-30.4}\,{\rm keV},\nonumber \\
\Gamma^{\rm NLO}_{Z \to B_c^*(2^3S_1)+b\bar{c}X} &=& 74.0^{+25.0}_{-31.3}\,{\rm keV}.
\label{eq.uncert4}
\end{eqnarray}

\end{widetext}

\subsection{PMC scale setting and results}

The sizable residual scale dependence motivates the use of the PMC
\cite{Brodsky:2011ta,Brodsky:2012rj,Mojaza:2012mf,Brodsky:2013vpa}.
Following the PMC method, we write the
integrated width at NLO as
\begin{equation}
\Gamma_{\rm NLO}=C_0\alpha_s^2(\mu_R)
 \left[1+(c_1+c_2n_f)\frac{\alpha_s(\mu_R)}{\pi}\right],
\label{eq.pmcdecomp}
\end{equation}
where $C_0$ is the LO coefficient, $c_1$ and $c_2$ are dimensionless
NLO coefficients, and $n_f$ is the number of active quark flavors. With $\beta_0=11-2n_f/3$, the
NLO coefficient becomes
\begin{equation}
c_1+c_2n_f=\left(c_1+\frac{33}{2}c_2\right)
 -\frac{3}{2}c_2\beta_0.
\label{eq.flavordecomp}
\end{equation}
The combination $c_1+33c_2/2$ is the conformal NLO coefficient.
The term proportional to $\beta_0$ is absorbed into the running
coupling using the renormalization group equation, giving
\begin{align}
Q_{\rm PMC}&=\mu_R\exp(3c_2/2),\nonumber\\
\Gamma_{\rm NLO}^{\rm PMC}
 &=C_0\alpha_s^2(Q_{\rm PMC})\nonumber\\
 &\quad\times\left[1+\left(c_1+\frac{33}{2}c_2\right)
 \frac{\alpha_s(Q_{\rm PMC})}{\pi}\right].
\label{eq.pmc}
\end{align}
The scale dependence of $c_2$ compensates the initial choice of
$\mu_R$ in $Q_{\rm PMC}$ at this order, fixing the scale from the
calculated NLO coefficient.

\begin{table}[h]
\centering
\caption{NLO decay widths (in keV) for
$Z\to B_c^{(*)}+b+\bar c+X$ using PMC scale setting.
The uncertainties arise from heavy-quark masses and wave functions
at the origin.}
\label{tb.pmc}

\renewcommand{\arraystretch}{1.4}
\begin{tabular}{c c}
\toprule
State & $\Gamma_{\rm PMC}$ (keV)\\ \colrule
$B_c(1^1S_0)$   & $119.7^{+38.3}_{-23.4}$\\
$B_c(2^1S_0)$   & $71.7^{+20.7}_{-20.3}$\\
$B_c^*(1^3S_1)$ & $138.0^{+50.6}_{-32.0}$\\
$B_c^*(2^3S_1)$ & $82.6^{+28.0}_{-25.3}$\\
\botrule
\end{tabular}
\end{table}

For the central inputs, we obtain
$Q_{\rm PMC}=5.0\,{\rm GeV}$ and $\alpha_s(Q_{\rm PMC})=0.215$
for $B_c$, and $Q_{\rm PMC}=2.9\,{\rm GeV}$ and
$\alpha_s(Q_{\rm PMC})=0.264$ for $B_c^*$.
The corresponding couplings are evaluated with $n_f=5$ and $n_f=4$,
respectively, using Eq.~(\ref{eq.alphas}) and the QCD scale parameters
specified above. Within each spin channel, the $1S$ and $2S$ states
share the same PMC scale because the radial wave function supplies
an overall normalization at the nonrelativistic order considered.

The PMC widths are given in Table~\ref{tb.pmc}. For the ground states,
the central values are $119.7\,{\rm keV}$ for $B_c$ and
$138.0\,{\rm keV}$ for $B_c^*$, which are $19.3\%$ and $11.6\%$
larger than the conventional NLO results at $\mu_R=2m_b$,
$100.4\,{\rm keV}$ and $123.7\,{\rm keV}$, respectively.
Both PMC predictions lie within the ranges spanned by the three
conventional scale choices in Tables~\ref{tb.widthBc} and
\ref{tb.widthBc*}. The channel-dependent shifts reflect the
reorganization of the NLO series through the absorption of its
$\beta_0$ term into the coupling.
The $2S$ widths are approximately $60\%$ of their $1S$ counterparts,
as set by the ratio of the central squared radial wave functions,
$0.983/1.642$.

With the renormalization scale fixed by the PMC prescription, the
uncertainties quoted in Table~\ref{tb.pmc} come from the heavy-quark
masses and wave functions at the origin. These input uncertainties
remain sizable. The quoted errors do not include contributions from
uncalculated higher orders or relativistic corrections.

\subsection{Differential decay widths}
\begin{figure}[!htbp]
    \centering
    \includegraphics[width=0.9\linewidth]{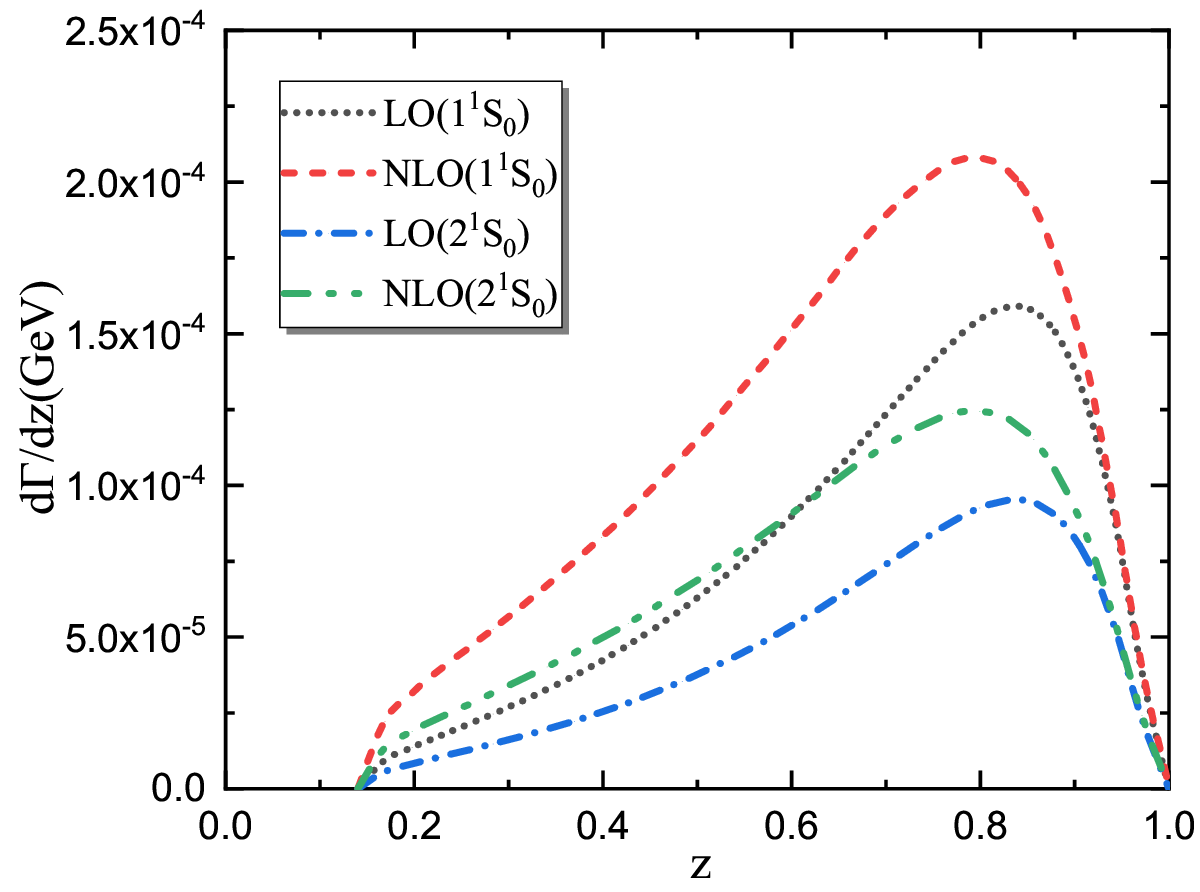}
    \caption{The LO and NLO differential decay widths $d\Gamma/dz$ for $Z \to B_c+b+\bar{c}+X$. }
    \label{fig:gammaz1s0}
\end{figure}

\begin{figure}[!htbp]
    \centering
    \includegraphics[width=0.9\linewidth]{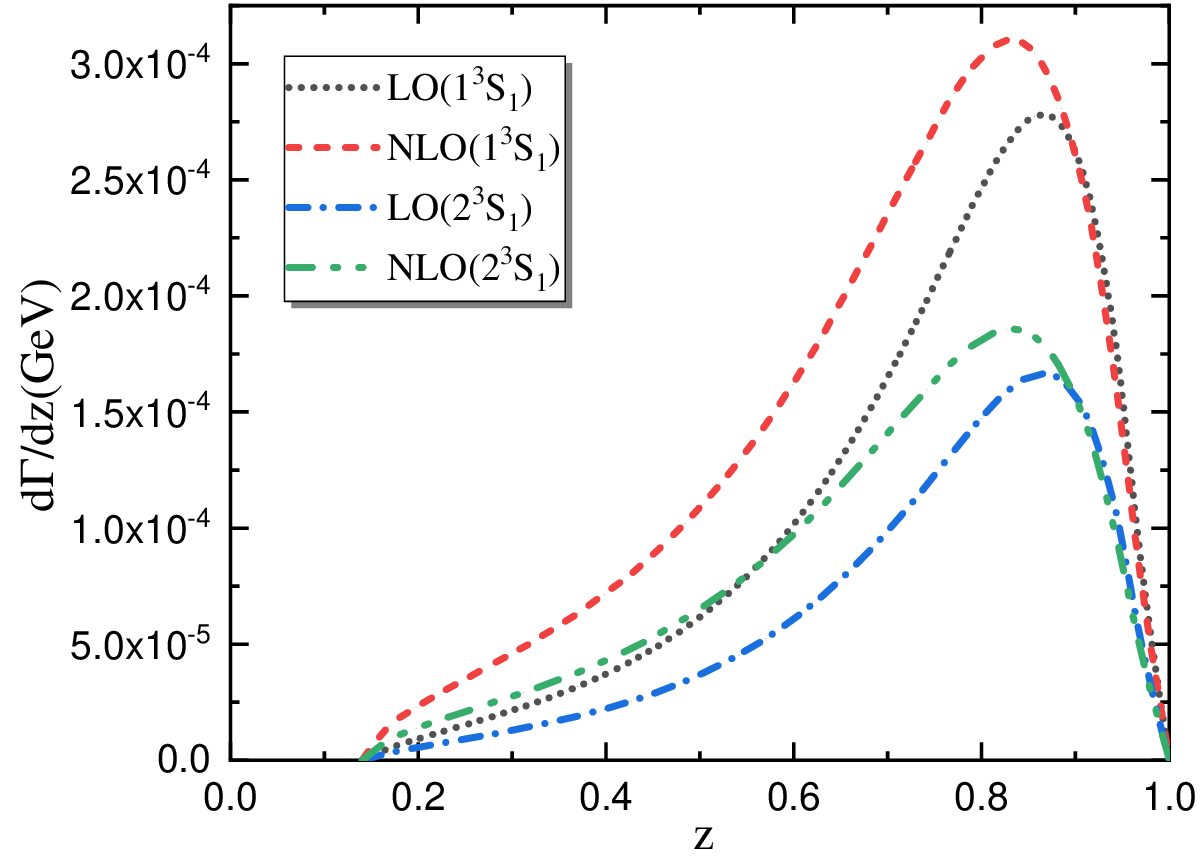}
    \caption{The LO and NLO differential decay widths $d\Gamma/dz$ for $Z \to B_c^*+b+\bar{c}+X$.}
    \label{fig:gammaz3s1}
\end{figure}

\begin{figure}[!htbp]
    \centering
    \includegraphics[width=0.9\linewidth]{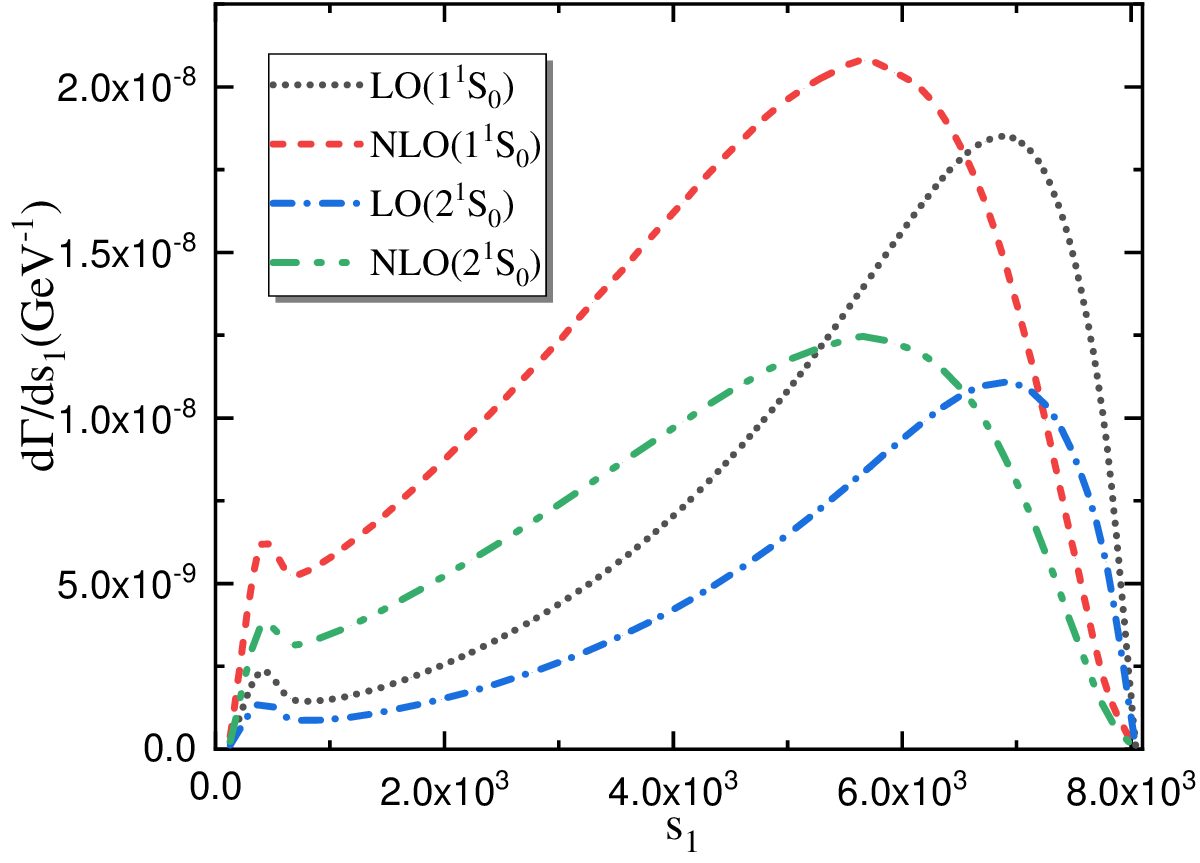}
    \caption{The LO and NLO differential decay widths $d\Gamma/d s_{1}$ for $Z \to B_c+b+\bar{c}+X$. The variable $s_1$ is in $\mathrm{GeV}^2$ and $d\Gamma/ds_1$ in $\mathrm{GeV}^{-1}$.}
    \label{gammas11s0}
\end{figure}

\begin{figure}[!htbp]
    \centering
    \includegraphics[width=0.9\linewidth]{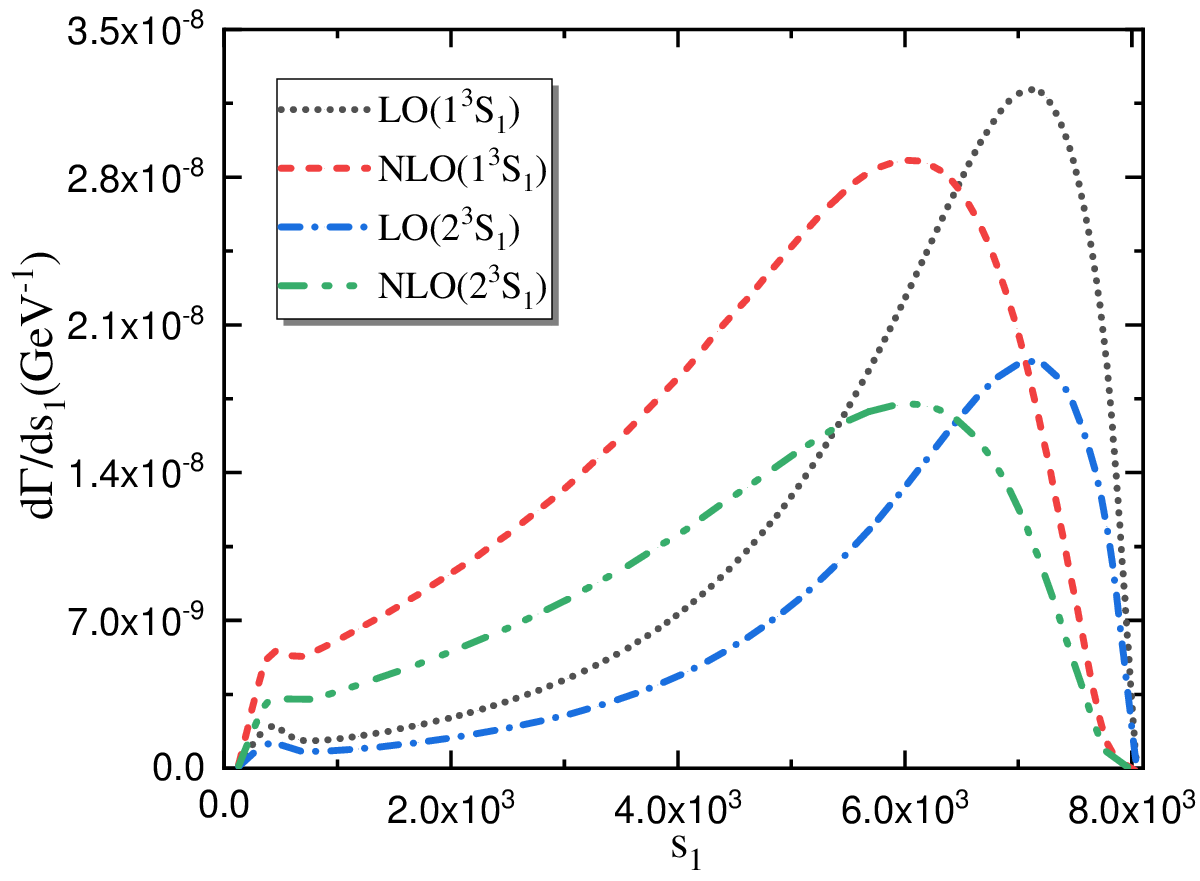}
    \caption{The LO and NLO differential decay widths $d\Gamma/d s_{1}$ for $Z \to B_c^*+b+\bar{c}+X$. The variable $s_1$ is in $\mathrm{GeV}^2$ and $d\Gamma/ds_1$ in $\mathrm{GeV}^{-1}$.}
    \label{gammas13s1}
\end{figure}

\begin{figure}[!htbp]
    \centering
    \includegraphics[width=0.9\linewidth]{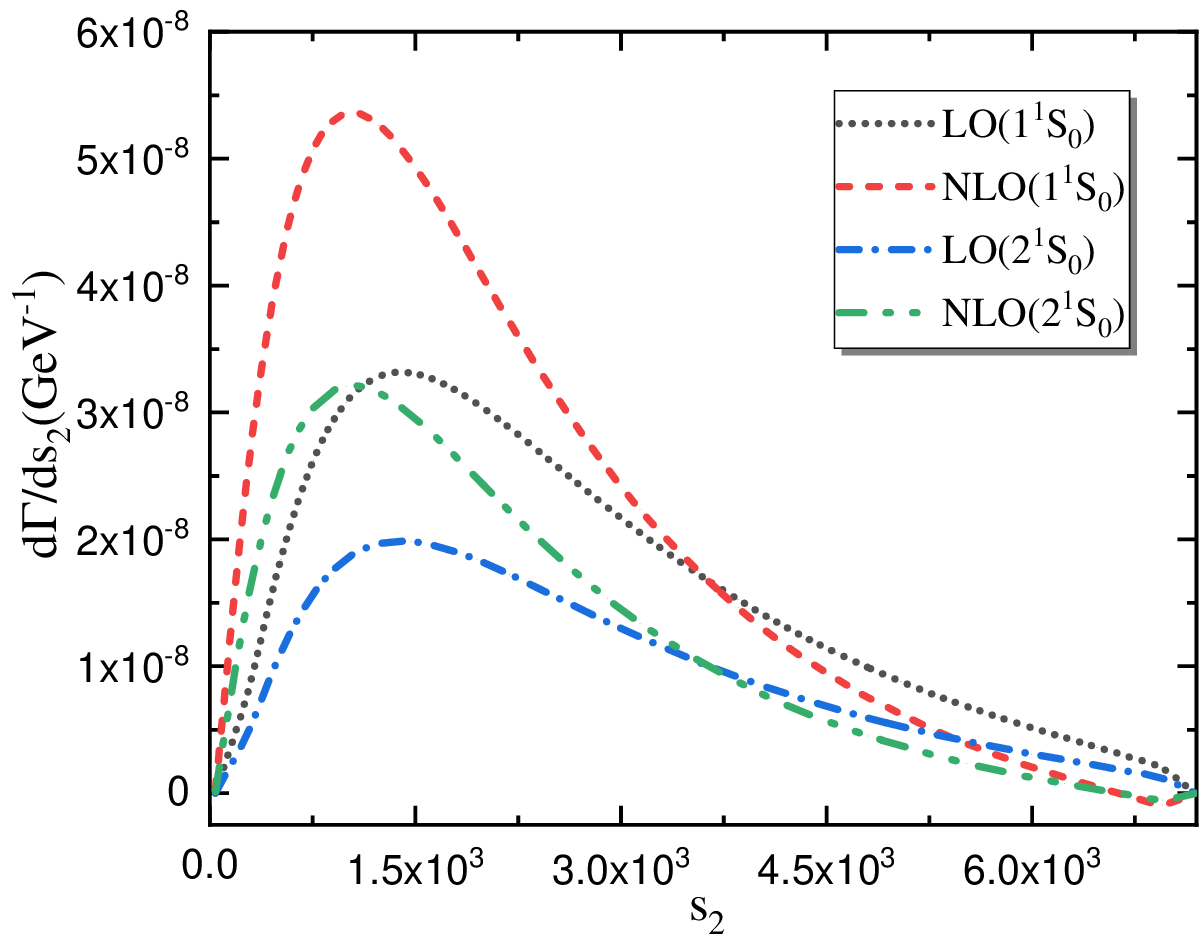}
    \caption{The LO and NLO differential decay widths $d\Gamma/d s_{2}$ for $Z \to B_c+b+\bar{c}+X$. The variable $s_2$ is in $\mathrm{GeV}^2$ and $d\Gamma/ds_2$ in $\mathrm{GeV}^{-1}$.}
    \label{gammas21s0}
\end{figure}

\begin{figure}[!htbp]
    \centering
    \includegraphics[width=0.9\linewidth]{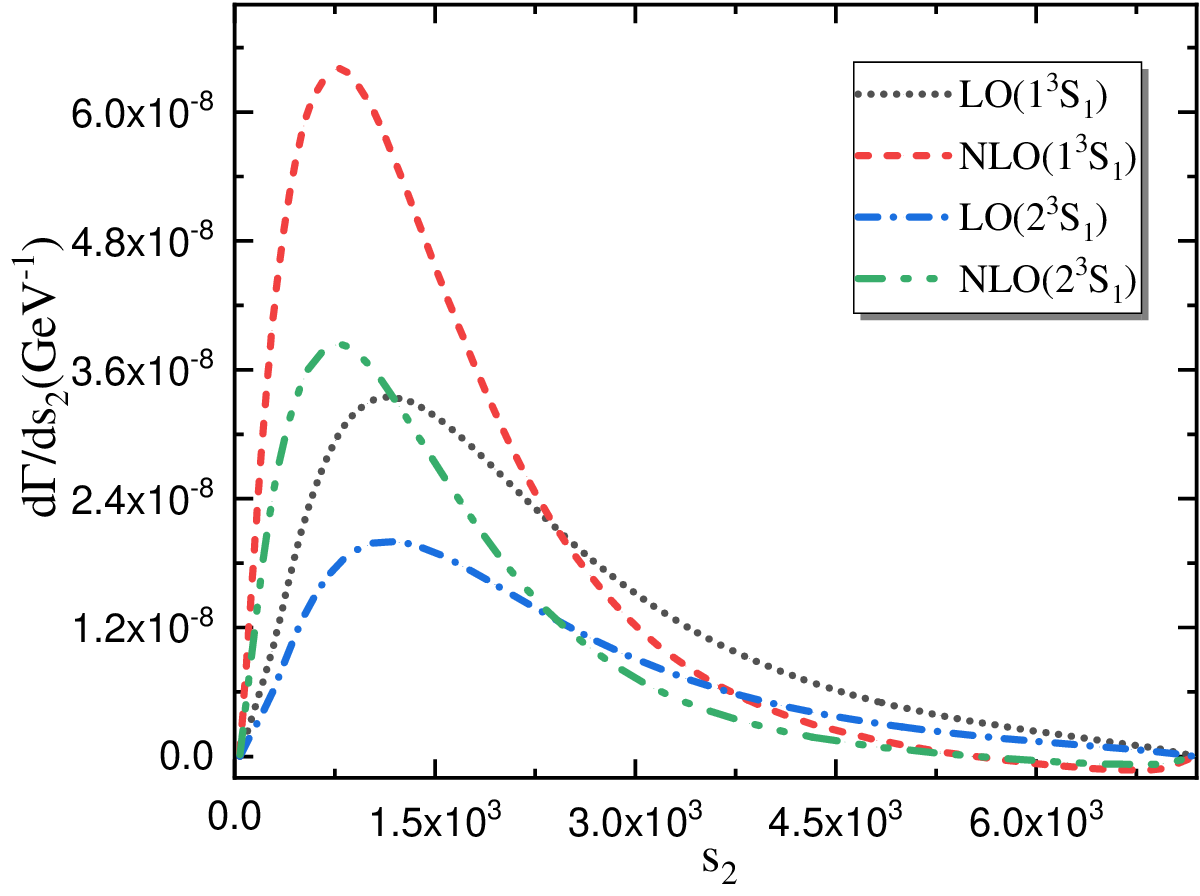}
    \caption{The LO and NLO differential decay widths $d\Gamma/d s_{2}$ for $Z \to B_c^*+b+\bar{c}+X$. The variable $s_2$ is in $\mathrm{GeV}^2$ and $d\Gamma/ds_2$ in $\mathrm{GeV}^{-1}$.}
    \label{gammas23s1}
\end{figure}

\begin{figure}[!htbp]
    \centering
    \includegraphics[width=0.9\linewidth]{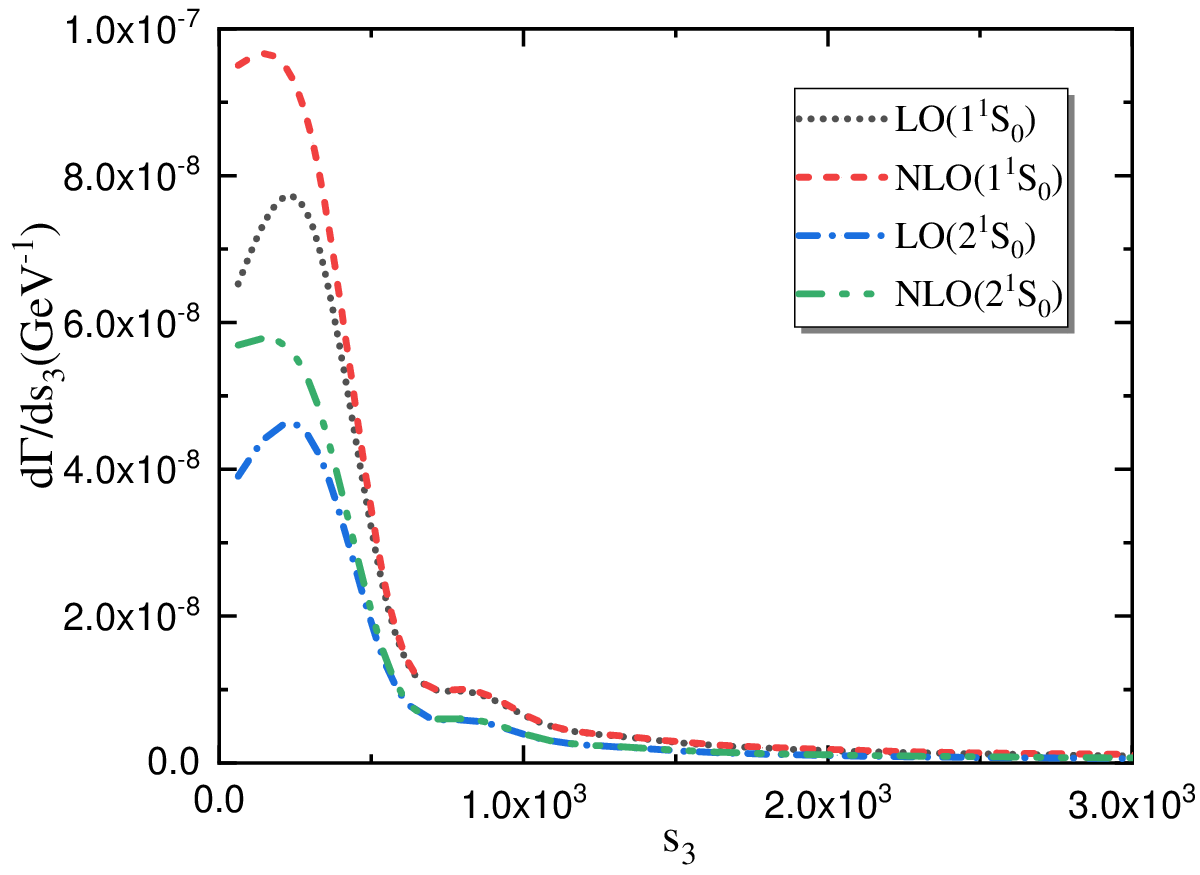}
    \caption{The LO and NLO differential decay widths $d\Gamma/d s_{3}$ for $Z \to B_c+b+\bar{c}+X$. The variable $s_3$ is in $\mathrm{GeV}^2$ and $d\Gamma/ds_3$ in $\mathrm{GeV}^{-1}$. }
    \label{gammas31s0}
\end{figure}

\begin{figure}[!htbp]
    \centering
    \includegraphics[width=0.9\linewidth]{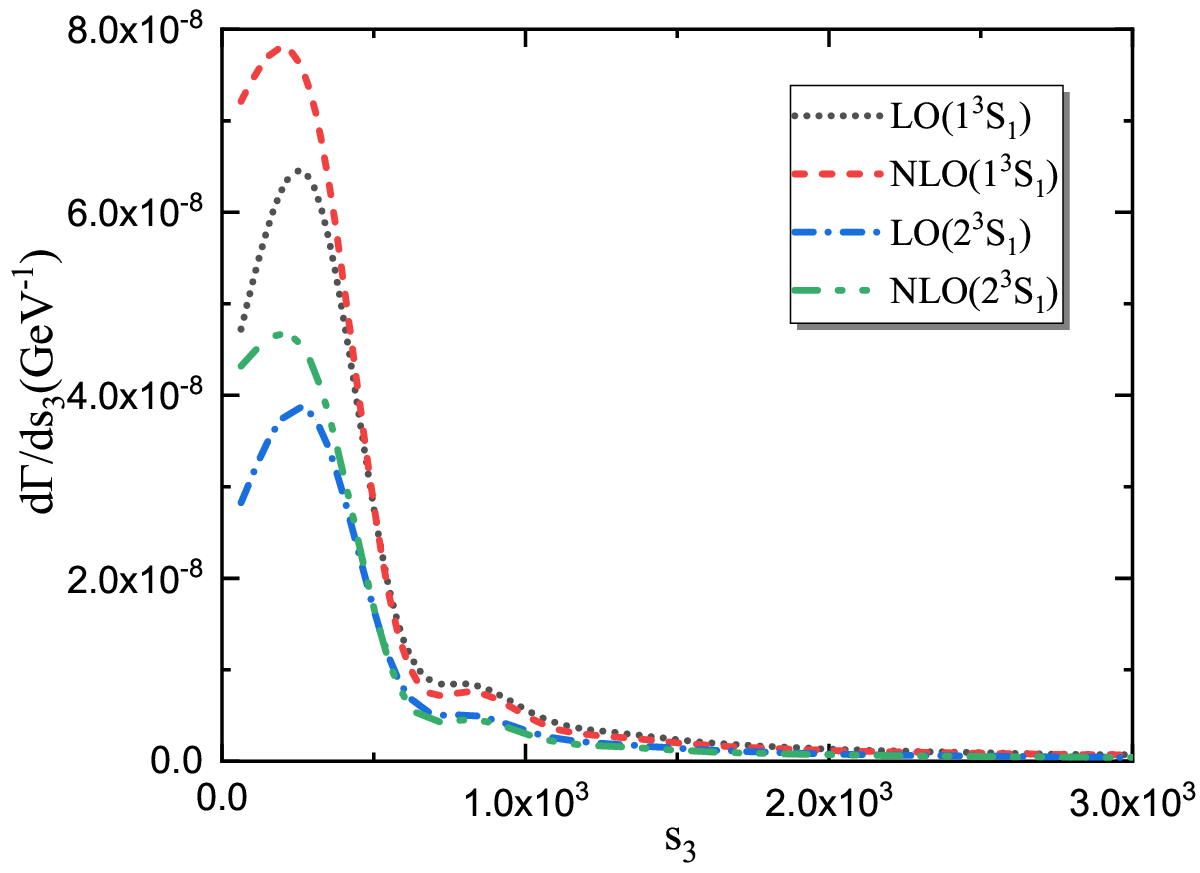}
    \caption{The LO and NLO differential decay widths $d\Gamma/d s_{3}$ for $Z \to B_c^*+b+\bar{c}+X$. The variable $s_3$ is in $\mathrm{GeV}^2$ and $d\Gamma/ds_3$ in $\mathrm{GeV}^{-1}$. }
    \label{gammas33s1}
\end{figure}

We consider the energy fraction and the three pair invariant masses
\begin{equation}
\begin{aligned}
z&=\frac{2p_0\cdot p_1}{m_Z^2},&
s_1&=(p_1+p_2)^2,\\
s_2&=(p_2+p_3)^2,&
s_3&=(p_1+p_3)^2.
\end{aligned}
\label{eq.observables}
\end{equation}
These definitions apply both to the Born channel and to real emission.
The gluon momentum $p_4$ is not recombined with either open heavy quark.
Thus $d\Gamma/dz$ has units of energy, whereas $d\Gamma/ds_i$ has
units of inverse energy. The figures use GeV and
$\mathrm{GeV}^{-1}$, respectively.

Figures~\ref{fig:gammaz1s0} and \ref{fig:gammaz3s1} show that the NLO
terms enhance the small- and intermediate-$z$ regions and modify the
large-$z$ behavior. The change contains both virtual corrections and
real-emission recoil; it cannot be attributed solely to the emitted
gluon. Within the common-mass approximation, the corresponding $2S$
spectra follow the normalization relation in Eq.~(\ref{eq.radial}).

The pair invariant-mass distributions in
Figs.~\ref{gammas11s0}--\ref{gammas33s1} provide complementary
information on the two heavy-quark fragmentation channels.
In $\bar b\to H+\bar c$, the meson and the accompanying $\bar c$
are produced approximately along the same direction, while the $b$
quark recoils against them. This configuration favors small $s_3$
and relatively large $s_1$. For $c\to H+b$, the meson is instead
correlated with the $b$ quark, while the $\bar c$ recoils, favoring
small $s_1$ and relatively large $s_3$.

The coexistence of these configurations provides a qualitative
explanation for the two separated enhancements in the $s_1$ spectra:
the low-mass enhancement is associated with $c$ fragmentation,
whereas the high-mass peak is associated with $\bar b$ fragmentation.
The more pronounced high-$s_1$ peak is consistent with the larger
$\bar b$-fragmentation contribution \cite{Braaten:1993jn}.
The displayed $s_3$ spectra emphasize the low-mass enhancement
characteristic of $\bar b$ fragmentation.
The NLO corrections enhance the low-$s_1$ and low-$s_3$ regions
and shift the dominant $s_1$ peak toward smaller invariant masses.
These correlations between the meson and the accompanying heavy
quarks provide useful kinematic signatures of the production
topologies.

\section{Summary}
\label{secSum}

We have calculated the NLO QCD corrections to
$Z\to B_c^{(*)}+b+\bar c+X$ within the color-singlet NRQCD framework
at leading order in the heavy-quark relative velocity. The analysis
includes both the $1S$ and $2S$ pseudoscalar and vector states and
provides integrated widths and differential distributions with both
bottom- and charm-quark production topologies included.

For the central inputs and $\mu_R=2m_b$, the $K$ factors are
$1.52$ for $B_c$ and $1.33$ for $B_c^*$, corresponding to
enhancements of $51.8\%$ and $33.2\%$ over the LO widths.
The radiative corrections thus have a sizable and spin-dependent
effect on the production rates. The NLO widths exhibit a smaller
variation across the three conventional scale choices than the LO
widths. We have also quantified the dependence on the heavy-quark
masses and radial wave functions at the origin.

Applying the PMC, we absorb the $\beta_0$-dependent NLO term into
the running coupling and obtain $Q_{\rm PMC}=5.0\,{\rm GeV}$
for $B_c$ and $Q_{\rm PMC}=2.9\,{\rm GeV}$ for $B_c^*$.
These scales are determined by the calculated NLO coefficients and
are independent of the initial renormalization scale at this order.
The corresponding $(1S,2S)$ widths are
$(119.7,71.7)\,{\rm keV}$ for $B_c$ and
$(138.0,82.6)\,{\rm keV}$ for $B_c^*$.
With the central wave-function inputs, the $2S$ widths are about
$60\%$ of their $1S$ counterparts in the common-mass approximation,
indicating a substantial direct production rate for radially excited
states.

The differential results show NLO enhancements at small and
intermediate $z$ and in the low-$s_1$ and low-$s_3$ regions,
together with a shift of the dominant $s_1$ peak toward smaller
invariant masses. The pair invariant masses probe complementary
correlations between the meson and the accompanying heavy quarks:
$\bar b$ fragmentation favors small $s_3$, whereas $c$ fragmentation
favors small $s_1$. This fragmentation picture provides a qualitative
interpretation of the calculated spectra.

The integrated widths and kinematic distributions provide NLO
benchmarks for studies of ground-state and radially excited $B_c$
production at high-statistics $Z$ factories. The excited-state widths
also supply production inputs for analyses of feed-down to the
ground state.

\begin{acknowledgments}
This work is supported in part by the National Natural Science Foundation of China under Grants No.~12205180, No.~12565014, and No.~12675111. The work of Jun Zeng is also supported by the Talent Research Startup Foundation of Hainan Normal University: HSZK-KYQD-202523, the Opening Foundation of Shanghai Key Laboratory of Particle Physics and Cosmology under Grant No. 22DZ2229013-5, and the Hainan Provincial Natural Science Foundation of China under Grant No. 126MS0134.
\end{acknowledgments}

\end{document}